# Electronic Structures, Born Effective Charges and Spontaneous Polarization in Magnetoelectric Gallium Ferrite

Amritendu Roy[1], Somdutta Mukherjee[2], Rajeev Gupta[2,3], Sushil Auluck[4], Rajendra Prasad[2] and Ashish Garg[1*]

[1] Department of Materials Science & Engineering, Indian Institute of Technology, Kanpur - 208016, India
[2] Department of Physics, Indian Institute of Technology, Kanpur - 208016, India
[3] Materials Science Programme, Indian Institute of Technology, Kanpur - 208016, India
[4] National Physical Laboratory, Dr. K. S. Krishnan Marg, New Delhi-110012, India

**ABSTRACT**

We present a theoretical study of the structure-property correlation in gallium ferrite, based on the first principles calculations followed by a subsequent comparison with the experiments. Local spin density approximation (LSDA+U) of the density functional theory has been used to calculate the ground state structure, electronic band structure, density of states and Born effective charges. Calculations reveal that the ground state structure is orthorhombic $Pc2_1n$ having A-type antiferromagnetic spin configuration, with lattice parameters matching well with those obtained experimentally. Plots of partial density of states of constituent ions exhibit noticeable hybridization of Fe *3d*, Ga *4s*, Ga *4p* and O *2p* states. However, the calculated charge density and electron localization function show largely ionic character of the Ga/Fe-O bonds which is also supported by lack of any significant anomaly in the calculated Born effective charges with respect to the corresponding nominal ionic charges. The calculations show a spontaneous polarization of ~ 59 μC/cm$^2$ along *b*-axis which is largely due to asymmetrically placed Ga1, Fe1, O1, O2 and O6 ions.

---

[*] Corresponding author, Tel: +91-512-2597904; FAX - +91-512-2597505, E-mail: ashishg@iitk.ac.in

# I. INTRODUCTION

Gallium ferrite (GaFeO$_3$ or GFO) is a piezoelectric and a ferrimagnet with its magnetic transition temperature close to the room temperature (RT).[1, 2] The transition temperature is affected largely by the Fe:Ga ratio within the single phase region (0.67≤ Fe/Ga ≤ 1.86) [3] and can be tuned to the values above RT. [[1, 3-5] As a result, accompanied by a good piezoelectric response,[6] compositionally modulated GFO is an exciting room temperature magnetoelectric material. Initial structural studies on this compound predicted the structure to be orthorhombic with *Pc2$_1$n* symmetry [4, 7, 8], confirmed subsequently by recent studies using neutron [1, 2, 9, 10] and x-ray diffraction [1, 3, 10, 11] investigations made on both powder and single crystals over a wide temperature range (4K-700K). The orthorhombic unit-cell comprises of eight formula units and the RT lattice parameters are: *a* = 8.7512 Å, *b* = 9.3993 Å, *c* = 5.0806 Å.[8] The unit-cell contains two nonequivalent Ga and Fe sites and there are six nonequivalent O sites. While Ga2, Fe1 and Fe2 are octahedrally coordinated by oxygen, Ga1 has tetrahedral coordination.[1] However, experimental observations reveal considerable cation site disorder indicating partial occupancy of Ga and Fe sites by Fe and Ga ions, respectively. [1, 8] The cation site disorder is also believed to be responsible for observed ferrimagnetism in GFO.[1]. Although not much has been reported on the structural distortion in GFO, asymmetric nature of Ga1-O tetrahedron is believed to contribute to the piezoelectricity in GFO with its piezoelectric coefficient being almost double to that of quartz. [12]

Despite a series of experimental studies, theoretical work, especially first-principles based calculations on GFO, have not really progressed, presumably because of the complex crystal structure and partial site occupancies of the cations. The only report by Han *et al.* [13] emphasizes on the magnetic structure and spin-orbit coupling behavior using a linear combination of localized pseudoatomic orbitals (LCPAO). However, there are no reports on the theoretical understanding of the structure, bonding and Born effective charges of GFO which is crucial to elucidate the structural distortion, nature of bonds and resulting polarization in GFO. Here, we present a first-principles density functional theory based calculation of the ground state structure of GFO along with experimental determination of structural parameters of a polycrystalline sample at room temperature. The calculations confirm that the ground state structure of GFO is A-type antiferromagnetic. We find that the Ga/Fe-O bonds have largely ionic character with no anomaly in the magnitude of Born effective charges. The calculations indicate the presence of a large spontaneous polarization (P$_s$) in GFO with a magnitude of ~ 59 μC/cm$^2$ along its *b*-axis.

# II. CALCULATION AND EXPERIMENTAL DETAILS

Our entire calculation is based on the first-principles density functional theory. [14] Vienna ab-initio simulation package (VASP) [15, 16] was used with the projector augmented wave method (PAW) [17]. The Kohn-Sham equation [18] was solved using the local spin density approximation (LSDA+U) [19] with the Hubbard parameter, U = 5 eV, and the exchange interaction, J = 1 eV. LSDA+U has been found to be quite efficient in describing strongly correlated multiferroic systems [20, 21] in comparison to the conventional local density approximation (LDA) and generalized gradient approximation (GGA). We employed the simplified, rotationally invariant approach introduced by Dudarev.[22] The value of U was optimized such that the moments of the magnetic ions are satisfactorily described with respect to the experiment. [1] We also checked that small variation of U from the optimized value does not alter the structural stability.

The calculations are based on the stoichiometric GFO assuming no partial occupancies of the constituent ions. We included 3 valence electrons of Ga ($4s^24p^1$), 8 for Fe ($3d^74s^1$) and 6 for O ($2s^22p^4$) ions. A plane wave energy cut-off of 550 eV was used. The conjugate gradient algorithm [23] was used for the optimization of the structure. All the calculations were performed at 0 K. Structural optimization and calculation of the electronic band structure and density of states were carried out using a Monkhorst-Pack [24] 7×7×12 mesh. Born effective charges, and spontaneous polarization for the ground state structure were calculated using Berry phase method [25] with a 3×3×3 mesh. A comparison of some of the results of the 3×3×3 mesh agree quite well with those obtained using a denser k-mesh. We also repeated some of our calculations using the generalized gradient approximation (GGA+U) with the optimized version of Perdew-Burke-Ernzerhof functional for solids (PBEsol) [26] to check the consistency of LSDA+U calculations. The effect of Ga $3d$ semicore state was studied with LSDA+U and GGA+U methods using a different pseudopotential of Ga that includes 13 valence electrons ($3d^{10}4s^24p^1$), while keeping all other pseudopotentials same. The calculations were performed using a Monkhorst-Pack 3×3×3 mesh. We started our calculations with the experimental structural parameters obtained from the neutron diffraction spectra of crushed single crystals of GaFeO$_3$ obtained at 4 K.[1] In order to obtain the ground state structure, ionic positions, lattice parameters and unit-cell shape were sequentially relaxed in such a way that the pressure on the optimized structure is almost zero and the Hellmann-Feynman forces are less than 0.001 eV/Å.

Born effective charge (BEC) tensor of an atom $k$, is defined as:

$$Z^*_{k,\gamma\alpha} = V\frac{\delta P_\gamma}{\delta \tau_{k,\alpha}} = \frac{\delta F_{k,\alpha}}{\delta \xi_\gamma} = -\frac{\partial^2 E}{\partial \xi_\gamma \partial \tau_{k,\alpha}} \qquad (1)$$

where, $P_\gamma$ represents polarization induced by the periodic displacement $\tau_{k,\alpha}$ or by the force $F_{k,\alpha}$ induced by an electric field $\xi_\gamma$. $E$ is the total energy of the unit cell. In the present calculation we displaced each ion by a small but finite distance along the three right handed Cartesian axes (unit-cell parameters are along the Cartesian axes), one at a time and calculated the polarization. Change in polarization with respect to undistorted structure divided by the displacement gives the elements of Born charges in a particular direction for an ion.

To corroborate the calculations with the experimental data, we synthesized a polycrystalline GaFeO$_3$ (Fe:Ga – 1:1) sample using the conventional solid-state-reaction route by mixing β-Ga$_2$O$_3$ and α-Fe$_2$O$_3$ powders. Powder diffraction data of the sintered pellet was collected on a Philips X'Pert Pro MRD diffractometer using Cu $K\alpha$ radiation.

Further, Rietveld refinement of the data was done using the FULLPROF 2000 [27] package using orthorhombic $Pc2_1n$ symmetry of GFO.

## III. RESULTS AND DISCUSSION

**A. Structural Optimization: Ground state structure**

To determine the ground state structure as well as to elucidate the magnetic structure of GFO, we considered four possible antiferromagnetic spin configurations as shown in Fig. 1(a)-(d) *i.e.* AFM-1 (A-type), AFM-2 (C-type), AFM-3 (G-type) and AFM-4. It should be noted that AFM-4 represents a possible spin configuration which is different from the conventional A, C and G-type. In addition to the above, we also considered other possible spin configurations which would ensure antiferromagnetism in GFO, but were found to be equivalent to one of the above shown in Fig. 1(a)-(d). While previous reports confirm the ground state structure of GFO to be antiferromagnetic [13], there is no discussion on the possible antiferromagnetic configurations. The results of total energy calculations of the four structures show that while energies of AFM-3 and AFM-4 structures are maximum (947.202 meV/unit-cell and 839.823 meV/unit-cell, respectively higher than AFM-1 structure); AFM-2 falls in the intermediate range with AFM-1 having the lowest energy. Hence, the stability of different spin configurations can, be ordered as: AFM-1 > AFM-2 > AFM-4 > AFM-3. On this basis, we can conclude that AFM-1 spin configuration is the most favored configuration in $Pc2_1n$ symmetry of GFO in the ground state. Hence, all further calculations were performed on AFM-1 structure.

Ground state crystal structure was determined by further relaxing the size, shape and ionic positions while maintaining AFM-1 spin configuration. The calculations show that the ground state structure retains the original $Pc2_1n$ symmetry observed experimentally at 298 K [8] and at 4 K [1] and also corroborated by our XRD data (shown in Fig. 2). A schematic representation of the ground state crystal structure is shown in the inset of Fig. 2. The calculated ground state lattice parameters, using LSDA+U, are: $a$ = 8.6717 Å, $b$ = 9.3027 Å and $c$ = 5.0403 Å which correspond well with our experimental data: $a$ = 8.7345 Å, $b$ = 9.3816 Å and $c$ = 5.0766 Å. Our calculation using GGA+U method yielded the ground state lattice parameters as follows: $a$ = 8.77119 Å, $b$ = 9.40936 Å and $c$ = 5.09811 Å. The calculated and experimentally determined lattice parameters are also in close agreement with the previously reported data: $a$ = 8.71932 Å, $b$ = 9.36838 Å and $c$ =5.06723 Å at 4 K [1], $a$ = 8.72569 Å, $b$ = 9.37209 Å and $c$ =5.07082 Å at 230 K [1], $a$ = 8.7512 Å, $b$ = 9.3993 Å and $c$ =5.0806 Å at 298 K [8]. Thus, the lattice parameters calculated using GGA+U and LSDA+U at 0 K are in good agreement with the experimental data obtained at 4 K [1], within a difference of ~ ±7 %. This difference can be attributed to the approximation schemes of LSDA and GGA. Moreover, it should be noted that the calculated ground state structure is perfectly ordered while the experimental structures may consist of partial cation site occupancies. Many first-principles calculations on Ga containing oxides include Ga *3d* as semicore states.[28] To investigate the effect of Ga *3d* semicore state, we also performed structural optimization of GFO using LSDA+U and GGA+U with a different pseudopotential of Ga that includes 13 valence electrons ($3d^{10}4s^24p^1$), while keeping all other pseudopotentials same. Structural optimization showed that the optimized lattice parameters are: $a$ = 8.642695 Å, $b$ = 9.271509 Å and $c$ = 5.023425 Å for LSDA+U and ($a$ = 8.836875 Å, $b$ = 9.479817 Å and $c$ = 5.136288 Å) for GGA+U. A comparison of these values with the experimental data as shown above, shows these to be even farther from the experimental data. While a comparison with the values calculated without considering Ga $3d$ semicore state shows that inclusion of Ga $3d$ semicore state slightly underestimates the lattice parameters in LSDA+U but overestimates in GGA+U. We, therefore, performed further calculations using the pseudopotential of Ga that includes 3 valence electrons ($4s^2 4p^1$) since it provides a better accuracy of the structural parameters.

The present experimentally determined ionic positions of stoichiometric GFO, along with the calculated ground state ionic positions are listed in Table 1 which shows that Fe1 and Fe2 ions lie on alternate planes parallel to the $ac$-plane. Since Fe1 and Fe2 have antiparallel spin configurations and are situated on alternate parallel planes, we conclude (see Fig 1) that the ground state magnetic structure of GFO is A-type antiferromagnetic. Fig. 2 (inset) also shows the coordination of the cations by oxygen: Ga1 is tetrahedrally coordinated while Ga2, Fe1 and Fe 2 are octahedrally coordinated by the surrounding oxygen ions.

From the positions of the ions in the calculated ground state structures and in the experimentally determined stoichiometric GFO at 298 K, we calculated the bond lengths of cations with neighboring oxygen ions. Table 2 consisting of calculated cation-oxygen and cation-cation bond lengths, shows a good agreement with the present and previous XRD [8] and neutron data [1]. Minor differences can be attributed to a number of factors, such as, temperature, site disorder and the limitation of the exchange correlation functionals used in our study. Using the bond length data from Table 2, we also calculated the structural distortions of the oxygen polyhedra. [29] The distortion can be quantified by determining the distortion index [30] which is defined as:

$$DI = \frac{1}{n}\sum_{i=1}^{n}\frac{(l_i - l_{av})}{l_{av}} \quad (2)$$

where $l_i$ is the bond length of $i^{th}$ coordinating ion and $l_{av}$ is the average bond length.

The calculations show DI values of Ga1-O tetrahedron is ~ 0.006 (at ground state for both LSDA+U and GGA+U) and 0.008 at 298 K and as a result, the effective anion co-ordination (~3.99 (LSDA+U and GGA+U), ~ 3.98 (expt.)) is almost identical to that of a regular tetrahedron *i.e.* 4. Here, the effective coordination number (ECoN) [31] is defined as:

$$ECoN = \sum_{i} \exp\left\{1 - \left(\frac{l_i}{l_{av}}\right)^6\right\} \quad (3)$$

In contrast, Ga2-O octahedron shows appreciable distortion (DI ~ 0.012 (LSDA+U), ~ 0.013 (GGA+U) and ~ 0.026 (expt.)) compared to a regular octahedron which is also reflected in a smaller co-ordination number of 5.93 (LSDA+U), 5.92 (GGA+U) and 5.75 (expt.) than the perfect octahedral co-ordination *i.e.* 6. This distortion is more significant in case of Fe1-O and Fe2-O octahedra with DI values of 0.056 (LSDA+U.), 0.057 (GGA+U) and 0.057 (expt.) and 0.063 (LSDA+U.), 0.062 (GGA+U) and, 0.059 (expt.), respectively while the corresponding average co-ordination numbers are 5.05 (LSDA+U.), 5.04 (GGA+U) and 4.74 (expt.) and 4.81 (LSDA+U.), 4.83 (GGA+U) and 4.92 (expt.), respectively. Thus, it is observed that for almost all the oxygen polyhedra, the cations are displaced from the center of the polyhedra. The significance of these distortions lies in imparting the non-centrosymmetry to the structure which results in the development of spontaneous polarization in GFO, as shown later in section III(C).

**B. Electronic Band Structure, Density of States and Bonding**

Fig. 3 shows the LSDA+U calculated electronic band structure along high symmetry directions and total density of states of GFO. The Fermi energy is fixed at 0 eV. The figure shows the plots of the band structure and total density of states demonstrating that the bands are spread over three major energy windows. The uppermost part of the valence band spreads over -7.73 eV to 0 eV. Above the Fermi level, the conduction band can again be divided into two parts: first part in the energy range from 1.77 eV to 2.45 eV while another part in the energy range from 3.0 eV to 16.83 eV (shown partially). The angular momentum character of the bands spread over different energy regions can be determined from the partial density of states (PDOS) of the constituent ions. PDOS of Fe1, Ga1 and O1 ions are shown in Fig. 4. As the nature of PDOS of the other ions is similar, these plots are not shown here. These figures show that the valence band (-7.73 eV to 0 eV) mainly consists of Fe *3d* and O *2p* states with significant amount of Ga *4s* and Ga *4p* characters also present in the lower energy side of this energy range. Beyond the Fermi level, a narrow energy band (1.77 eV to 2.45 eV) contains mainly Fe *3d* character. The highest energy window (3.0 eV to 16.83 eV) has contributions from Fe *3d*, Ga *4s*, Ga *4p* and O *2p* states. More importantly, PDOS demonstrate significant hybridization of Fe *3d*, Ga *4p* and O *2p* states throughout the uppermost part of the valence band. Such hybridization of transition metal *d* state and O *2p* state has been found to impart ferroelectricity in a number of perovskite oxides [32, 33] and can be of interest in GFO too.

As shown in Fig. 3, our LSDA+U calculations yielded a direct band gap ($E_g$) of ~2.0 eV (Γ- Γ) while GGA+U calculations showed a direct band gap of ~ 2.25 eV. Calculation of band structure using LSDA+U method with pseudopotential treating Ga *3d* as semicore state, did not reveal any noticeable change from that of our earlier calculation and a direct band gap ($E_g$) of ~ 1.98 eV (Γ- Γ) was obtained. However, experimental studies based on optical absorption spectra of GFO report a band gap of 2.7-3.0 eV.[34] The difference between calculated band gap and the experimental data is expected (due to underestimation of band gap by the LSDA and GGA methods) and is common in electronic structure calculation of oxides. [35, 36]

Moreover, PDOS data in Fig. 4 can also shed light on the bonding behavior in GFO, especially partial covalency of cation-anion bonds, which can be further correlated with the functional properties of GFO. From Fig. 4, we find that Fe *3d* and O *2p* states are significantly hybridized in the uppermost part of the valence band in GFO. For a detailed analysis, we have plotted the charge density distribution calculated using LSDA+U, on three principal planes of the unit cell as shown in Fig. 5(a). The figure shows that although most of the charges are symmetrically distributed along the radius of the circles, indicating largely ionic nature of bonding, small amount of covalency is shown by minor asymmetry of the charges around O ions connected to Fe1, Fe2, Ga1 and Ga2 ions.

However, nature of binding interaction as determined from the charge density distribution alone is not conclusive. We, therefore, utilized electron localization function (ELF) which provides a measure of the local influence of the Pauli repulsion on the behavior of the electrons and allows the mapping of core, bonding and nonbonding regions of the crystal in real space. Thus ELF can be used as a tool to differentiate the nature of different types of bonds. [37] A large value of ELF function indicates a region of small Pauli repulsion, in other words, space with anti-parallel spin configuration while the position with maximum ELF value has signature of electrons pair. [37] Fig. 5(b) and (c) show the ELF distribution in three principal planes and in the entire unit cell of GFO, respectively, calculated by LSDA+U method. Fig. 5(b) also depicts the maximum ELF value at O sites and small values at Fe and Ga sites indicating charge transfer interaction from Fe/Ga to O sites. Comparing Fig. 5(a) and (b), we find that almost complete charge transfer takes place between Fe2 and O3 ions. Similar charge transfer, albeit to a lesser extent, is also observed between Fe1 and O1, O2 ions. Thus we can conclude that the Fe-O bonds in GFO are mostly ionic. In contrast, polarization of ELF from O sites toward other O sites and finite value of ELF between O and Ga1 (Fig. 5(b)) indicate some degree of covalent characteristics. Similar feature is expected for Ga2-O bonds as shown in Fig. 5(c). Therefore, from the charge density and electron localization function plots, we can assert that Ga/Fe- O bonds in GFO are largely of ionic character. The ionicity is greater for Fe-O bonds, while some degree of hybridization is observed in Ga-O bonds indicating covalency.

**C. Born Effective Charge and Spontaneous Polarization**

The nature of bonding can further be correlated with the Born effective charges ($Z^*$), defined in section II. These charges are important quantities in elucidating the physical understanding of piezoelectric and ferroelectric properties since they describe the coupling between lattice displacements and the electric field. Born charges are also indicators of long range Coulomb interactions whose competition with the short range forces leads to the ferroelectric transition. Previous studies on many perovskite ferroelectric show anomalously large Born charges for some of the ions [32, 33] which are often explained as manifestation of strong covalent character of bonds between the specific ions. In GFO, from the charge density and ELF plots, we have observed that charge sharing between the Ga/Fe and O ions in cation-oxygen bonds is not significant in comparison to conventional perovskite ferroelectrics. [32, 33] On the other hand, from the structural data we find that the cation-oxygen polyhedra are highly distorted. Since ferroelectric and piezoelectric responses are combined manifestations of structural distortions and effective charges of the constituent ions [38], it is imperative to calculate the Born effective charges of the constituent ions in GFO. Such a calculation would help to elucidate the nature of cation-oxygen bonds and the origin of polarization in the material.

In the present work, we have calculated the Born effective charge tensors of nonequivalent ions in $Pc2_1n$ structure of GFO by slightly displacing each ion, one at a time, along three axes of the Cartesian co-ordinates and then calculating the resulting difference in polarization, using Berry phase method. [25] We used LSDA+U technique for this calculation. Table 3 lists the three diagonal elements of the Born effective charge tensors of each ion along with their nominal charges. Here, we observe that that Ga1 ion has elements of effective charge tensors close its nominal ionic charge and hence, it is concluded that all the bonds between Ga1 and surrounding O ions are primarily ionic in nature. On the other hand, Ga2 develops a maximum effective charge of 3.53, ~ 18 % higher with respect to its static charge of +3. In contrast, both Fe1 and Fe2 ions show much higher increase in the effective charges, 36 % and 27 % respectively, while oxygen ions show a maximum reduction of 39.5 % with respect to the nominal ionic charge. Interestingly, all these elements that have maximum change with respect to the respective static charges are along $z$-axis (except for Ga1). However, the direction of $P_s$ is along $y$-axis i.e. crystallographic $b$-direction. [1] Hence, unlike in most perovskite ferroelectrics [32, 33], the polarization in GFO is not due to large effective ionic charges. Instead, it is most likely to be caused by the structural distortion and noncentrosymmetry of the structure.

To compare our results on Born effective charges with the effective charges calculated by other methods, we calculated these charges on each ionic site using bond valence method in which bond valence charge (V) is defined as:

$$V = \sum_i v_i = \sum_i \exp(\frac{R_0 - R_i}{b}) \qquad (4)$$

where, $R_0$ is the ideal bond length for a bond with valence 1, $R_i$ is measured bond length and $b$ is an empirical constant. We have also estimated effective charge distribution [29] at different ionic sites based on the nominal ionic charges and polyhedra parameters. The results obtained from both methods are shown in Table 3. Though these calculations are in no way comparable to the *ab-initio* calculations, they are useful in getting a trend of the effective charges. The comparison shows that although the calculated Born effective charges using *ab-initio* method are larger than the effective charges calculated using bond valence method and charge distribution method, all the calculations of effective charges point toward the fact that the cation-oxygen bonds in GFO are largely ionic and substantiate the discussion in the preceding paragraph.

The Born effective charges can also be used to quantify the spontaneous polarization in GFO. Although previous studies [1, 12] indicate the direction of $P_s$ along [010]-direction, there is no conclusive experimental report on the value of $P_s$. Although Arima *et al* [1] predicted a $P_s$ ~ 2.5 µC/cm$^2$ based on the displacement of Fe ions from the center of FeO$_6$ octahedra, such point charge calculation does not provide a correct estimate since various other contributions to $P_s$ were neglected. As we see later, these other contributions are from the sources such as Ga1-O tetrahedra and Ga2-O octahedra, and more importantly, effective ionic charges. To compare, we have calculated $P_s$ of GFO in its ground state using both nominal ionic charges and calculated Born effective charges.

From the crystallography perspective, GFO having *Pc2₁n* space group, allows following symmetry operations to be performed: (i) *c*-operation, a glide translation along half the lattice vector of *c*-axis leading to (½–x, y, ½+z), (ii) *2₁* operation, 2-fold screw rotation around *b*-axis leading to (-x, ½+y, -z) and (iii) *n*-operation, a glide translation along half of the face-diagonal leading to (½+x, ½+y, ½-z). Here, we observe that the application of first and third operations (*c* and *n* respectively) on the atom positions does not put any constraint on the displacement and in turn polarization vector remains unrestricted. However, when 2₁ symmetry operator is applied i.e. when the cell is screw rotated by 180° about [010]-axis i.e. *b*-axis, it changes the crystal polarization from (Px, Py, Pz) to (-Px, Py, -Pz) as (x,y,z) becomes (-x,y,-z). This shows that the crystal polarization along *a*- and *c*-axis is equal to zero and is non-zero along b-axis. Further, using the Born effective charges from Table 3, we calculated the spontaneous polarization ($P_s$) as ~ 58.63 µC/cm$^2$ which is an order of magnitude larger than that predicted by Arima *et al.* [1]. Similar calculation using the nominal ionic charges yielded $P_s$ of ~ 30.53 µC/cm$^2$, almost half the value obtained using the Born effective charges. We, therefore, conclude that though the values of Born charges of the constituent ions are not anomalously large unlike some perovskite ferroelectrics [32, 33], they do seem to affect the spontaneous polarization response in GFO rather significantly.

We also calculated partial polarization in order to estimate the relative contribution of individual ions. A schematic of the partial polarization contributions from individual ions toward the total spontaneous polarization has been shown in Fig. 6. It was found that while the contribution from Ga1 is the largest, it is counter-balanced by the opposite contributions from Fe1, O1, O2 and O6. Interestingly, the structure data (Table 1 and Fig. 2) also shows that these ions are the most asymmetrically placed around the inversion center of symmetry while Ga2 and Fe2 cations maintain almost centrosymmetric configuration and contribute least to the total polarization. Therefore, we conclude that the spontaneous polarization in GFO is primarily contributed by the asymmetrically placed Ga1, Fe1, O1, O2 and O6 ions. However, at elevated temperatures, the site disordering between Fe1 and Ga1 sites is expected [1] which may substantially lower the spontaneous polarization. This should be of interest for further theoretical investigations incorporating the effect of disorder on calculations.

**CONCLUSIONS**

We have presented a theoretical study of the structure-property relationship in gallium ferrite (GFO), supported by the experimental data. First-principles density functional theory based calculations were performed to calculate the ground state structure of GFO. The calculations support an orthorhombic structure with *Pc2₁n* symmetry and A-type antiferromagnetic spin configuration in the ground state with calculated ground state lattice parameters, bond strength and bond angles agreeing well with the experimental results. While, the electronic density of states show hybridization among Fe *3d*, Ga *4s*,

Ga *4p* and O *2p* states, calculations of electronic charge density demonstrate almost symmetrical charge distribution on most of the major planes indicating an ionic nature of bonds. Calculation of the electron localization function further supported a largely ionic character of Fe-O bonds and a finite degree of hybridization among O, Ga1 and Ga2 ions. Moreover, lack of any significant anomaly in the Born effective charges with respect to the corresponding nominal ionic charges again emphasized towards ionic character of the bonds. Calculations also showed a spontaneous polarization of ~ 59 $\mu C/cm^2$ along *b*-direction *i.e.* [010]-axis, attributed primarily to the structural distortion.

## Acknowledgements

Authors thank Prof. M.K. Harbola (Physics, IIT Kanpur) for valuable discussions on the work and his suggestions. AR and SM thank Ministry of Human Resources, Government of India for the financial support.

**Tables**

Table 1 – Calculated ground state ionic positions of orthorhombic ($Pc2_1n$) GFO using LSDA+U and GGA+U along with Rietveld refined experimental data.

| Ion | LSDA+U | | | GGA+U | | | Experiment at 298 K | | |
|---|---|---|---|---|---|---|---|---|---|
| | X | y | z | x | Y | z | x | y | z |
| Ga1 (4a) | 0.15101 | 0.99844 | 0.17665 | 0.15125 | 0.99844 | 0.175969 | 0.15291 | 0.00000 | 0.17900 |
| Ga2 (4a) | 0.16068 | 0.30818 | 0.81637 | 0.16087 | 0.30817 | 0.81653 | 0.15902 | 0.30413 | 0.81446 |
| Fe1 (4a) | 0.15512 | 0.58224 | 0.18817 | 0.15477 | 0.58248 | 0.18690 | 0.15299 | 0.58079 | 0.20291 |
| Fe2 (4a) | 0.03075 | 0.79453 | 0.67380 | 0.03078 | 0.79453 | 0.67314 | 0.03197 | 0.79907 | 0.67050 |
| O1 (4a) | 0.32292 | 0.42757 | 0.98443 | 0.32260 | 0.42709 | 0.98386 | 0.32120 | 0.42638 | 0.98250 |
| O2 (4a) | 0.48576 | 0.43140 | 0.51922 | 0.48600 | 0.43128 | 0.51976 | 0.98915 | 0.43217 | 0.51623 |
| O3 (4a) | 0.99672 | 0.20019 | 0.65659 | 0.99694 | 0.20084 | 0.65734 | 0.99730 | 0.19794 | 0.66331 |
| O4 (4a) | 0.16218 | 0.19907 | 0.15803 | 0.16176 | 0.19902 | 0.15796 | 0.16015 | 0.19924 | 0.14523 |
| O5 (4a) | 0.16719 | 0.67266 | 0.84410 | 0.16752 | 0.67224 | 0.84306 | 0.15901 | 0.66492 | 0.84351 |
| O6 (4a) | 0.16636 | 0.93781 | 0.52144 | 0.16635 | 0.93800 | 0.52079 | 0.16260 | 0.94593 | 0.52414 |

Table 2 - Calculated bond lengths from the ground state ionic positions of orthorhombic ($Pc2_1n$) GFO along with experimental data from the present work and previously reported data (*: Present work, $: Ref. [1], #: Ref. [7] )

| Bond length (Å) | Theory | | Experimental Data | | | %Difference (LSDA+U-Experiment at 4 K) |
|---|---|---|---|---|---|---|
| | LSDA+U | GGA+U | 298 K* | 4 K$ | 298 K# | |
| Ga1-O2  | 1.849 | 1.869 | 1.853 | 1.844 | 1.851 | 0.27 |
| Ga1-O6  | 1.832 | 1.852 | 1.826 | 1.822 | 1.813 | 0.55 |
| Ga1-O6' | 1.854 | 1.873 | 1.863 | 1.836 | 1.867 | 0.98 |
| Ga1-O4  | 1.871 | 1.892 | 1.878 | 1.857 | 1.852 | 0.75 |
| Ga2-O3  | 1.918 | 1.935 | 1.891 | 1.892 | 1.927 | 1.37 |
| Ga2-O1  | 1.983 | 1.998 | 2.012 | 1.985 | 2.011 | -0.10 |
| Ga2-O2  | 1.993 | 2.019 | 2.041 | 2.006 | 2.054 | -0.65 |
| Ga2-O4  | 2.007 | 2.032 | 2.050 | 2.059 | 2.077 | -2.53 |
| Ga2-O4' | 1.999 | 2.021 | 1.946 | 1.996 | 2.037 | 0.15 |
| Ga2-O1' | 2.013 | 2.037 | 2.046 | 2.053 | 2.051 | -1.95 |
| Fe1-O1  | 2.082 | 2.114 | 2.041 | 2.064 | 2.058 | 0.87 |
| Fe1-O1' | 2.291 | 2.319 | 2.347 | 2.354 | 2.361 | -2.68 |
| Fe1-O2  | 2.046 | 2.068 | 2.094 | 2.074 | 2.06  | -1.35 |
| Fe1-O3  | 1.884 | 1.908 | 1.842 | 1.905 | 1.866 | -1.10 |
| Fe1-O5  | 1.923 | 1.943 | 1.957 | 1.918 | 1.936 | 0.26 |
| Fe1-O5' | 1.930 | 1.949 | 1.989 | 1.934 | 1.934 | -0.21 |
| Fe2-O1  | 2.326 | 2.352 | 2.328 | 2.324 | 2.354 | 0.09 |
| Fe2-O2  | 2.042 | 2.064 | 2.056 | 2.025 | 2.064 | 0.84 |
| Fe2-O4  | 2.075 | 2.098 | 2.137 | 2.131 | 2.093 | -2.63 |
| Fe2-O3  | 1.897 | 1.917 | 1.959 | 1.943 | 1.946 | -2.37 |
| Fe2-O5  | 1.850 | 1.874 | 1.894 | 1.875 | 1.872 | -1.33 |
| Fe2-O6  | 1.936 | 1.959 | 1.937 | 1.958 | 1.971 | -1.12 |
| Fe1-Fe2 | 3.201 | 3.240 | 3.164 | 3.201 | 3.234 | 0 |
| Ga1-Ga2 | 3.231 | 3.271 | 3.286 | 3.246 | - | -0.46 |
| Fe2-Ga2 | 3.062 | 3.100 | 3.102 | 3.089 | 3.007 | -0.87 |
| Fe1-Ga1 | 3.320 | 3.354 | 3.387 | 3.328 | - | -0.24 |
| Fe1-Ga2 | 3.165 | 3.198 | 3.123 | 3.216 | 3.121 | -1.59 |

Table 3- Diagonal elements of the Born effective charge tensors computed using Berry phase technique within LSDA+U. The bond valence charges (V) were calculated using bond length data based on the ground state structural parameters. Nominal ionic charges are also provided for comparison.

| Ion | Nominal ionic charge (e) | Z* (e) | | | Charge distribution (e) | Bond valence charge (e) |
|---|---|---|---|---|---|---|
| | | $Z_{xx}$ | $Z_{yy}$ | $Z_{zz}$ | | |
| Ga1 | 3 | 3.01 | 3.11 | 2.99 | 2.83 | 2.88 |
| Ga2 | 3 | 3.57 | 3.16 | 3.53 | 3.23 | 3.02 |
| Fe1 | 3 | 3.66 | 3.78 | 4.08 | 3.04 | 3.10 |
| Fe2 | 3 | 3.68 | 3.38 | 3.82 | 2.90 | 3.20 |
| O1 | -2 | -2.29 | -2.58 | -2.79 | -1.56 | - |
| O2 | -2 | -2.45 | -2.29 | -2.41 | -2.12 | - |
| O3 | -2 | -2.54 | -2.30 | -2.75 | -2.04 | - |
| O4 | -2 | -2.27 | -2.85 | -2.17 | -2.02 | - |
| O5 | -2 | -2.50 | -2.16 | -2.79 | -2.10 | - |
| O6 | -2 | -2.32 | -2.08 | -2.40 | -2.16 | - |

**List of Figures**

Fig. 1  Schematics of different antiferromagnetic spin configurations considered in the present calculations. The configurations are assigned as (a) AFM-1 (A-type), (b) AFM-2 (C-type), (c) AFM-3 (G-type) and (d) AFM-4 (Different variant).

Fig. 2  Rietveld refinement of room temperature XRD data of stoichiometric GFO. Inset shows schematic of the crystal structure of $GaFeO_3$ having orthorhombic $Pc2_1n$ symmetry.

Fig. 3  Electronic structures of orthorhombic ($Pc2_1n$) $GaFeO_3$ calculated using the LSDA+U method. Left panel shows plot of total density of states as a function of energy while right panel shows electronic band structure along high symmetry directions. The zero in the energy axis is set at the highest occupied level.

Fig. 4  PDOS plots of Ga1 4s and 4p states, Fe1 *3d* state and O1 *2s* and *2p* states calculated using the LSDA+U method. The vertical blue line indicates the Fermi level.

Fig. 5  Plots of (a) charge density along three principal planes of $GaFeO_3$ unit cell, (b) electron localization function calculated using the LSDA+U method along three principal planes of $GaFeO_3$ unit cell keeping the area of the planes in accordance with the respective lattice parameters and (c) 3-D image of electron localization function distribution in the $GaFeO_3$ unit cell.

Fig. 6  Schematic diagram showing partial polarization of individual ions along crystallographic *b*-direction. The strength and direction of polarization is depicted by the size and direction of the arrows.

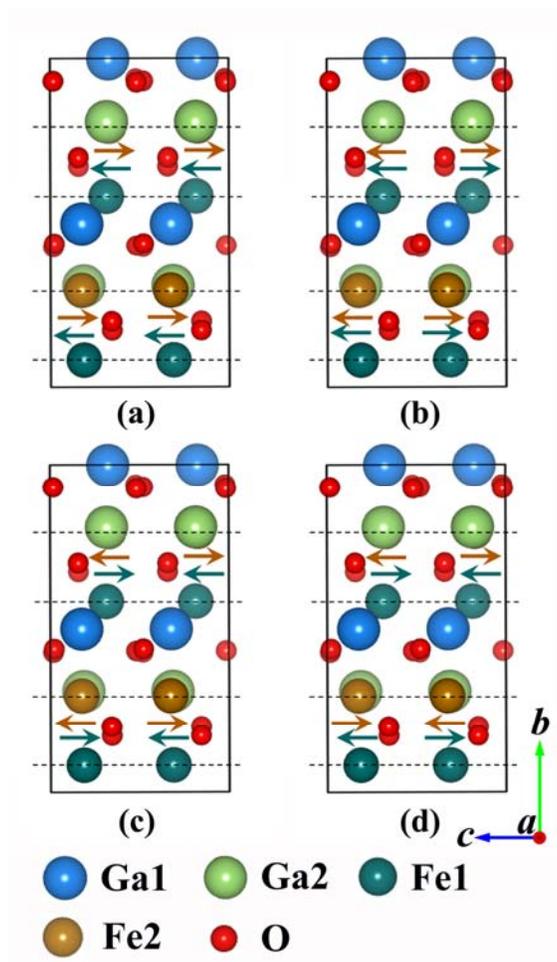

Fig. 1. Roy *et al*,

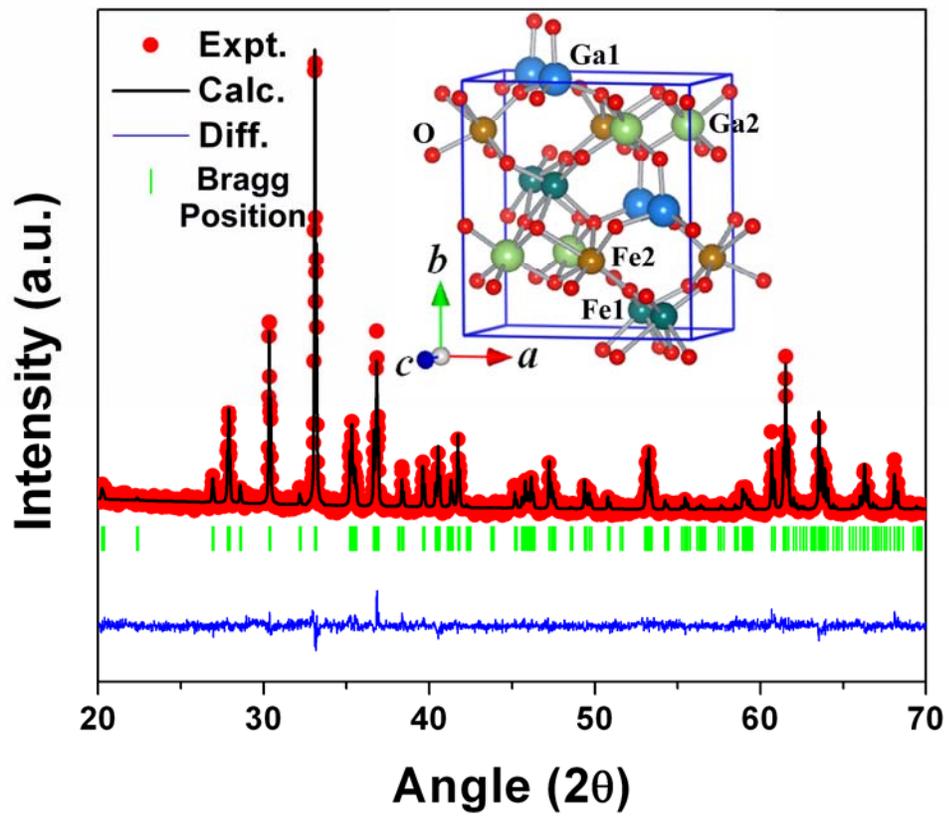

Fig. 2. Roy *et al*,

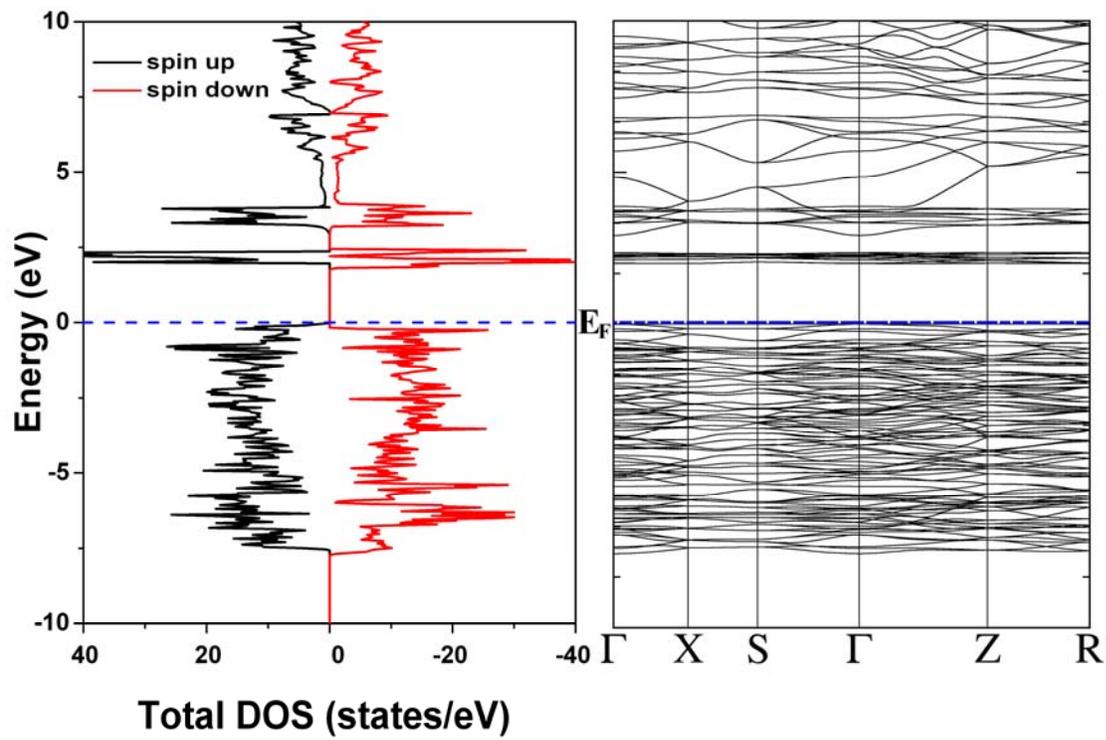

Fig. 3. Roy *et al*,

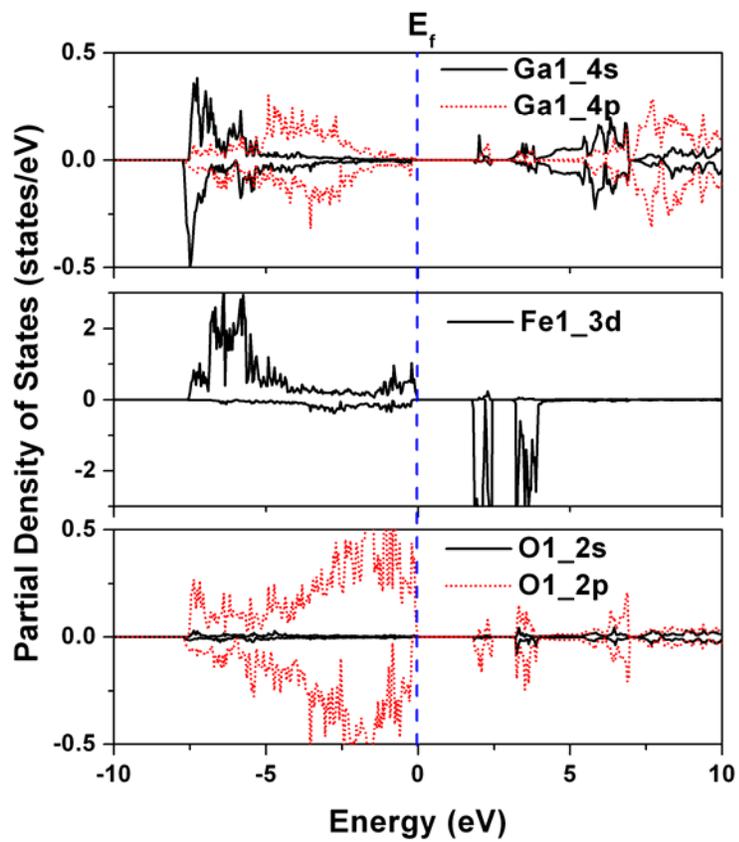

Fig. 4. Roy *et al*,

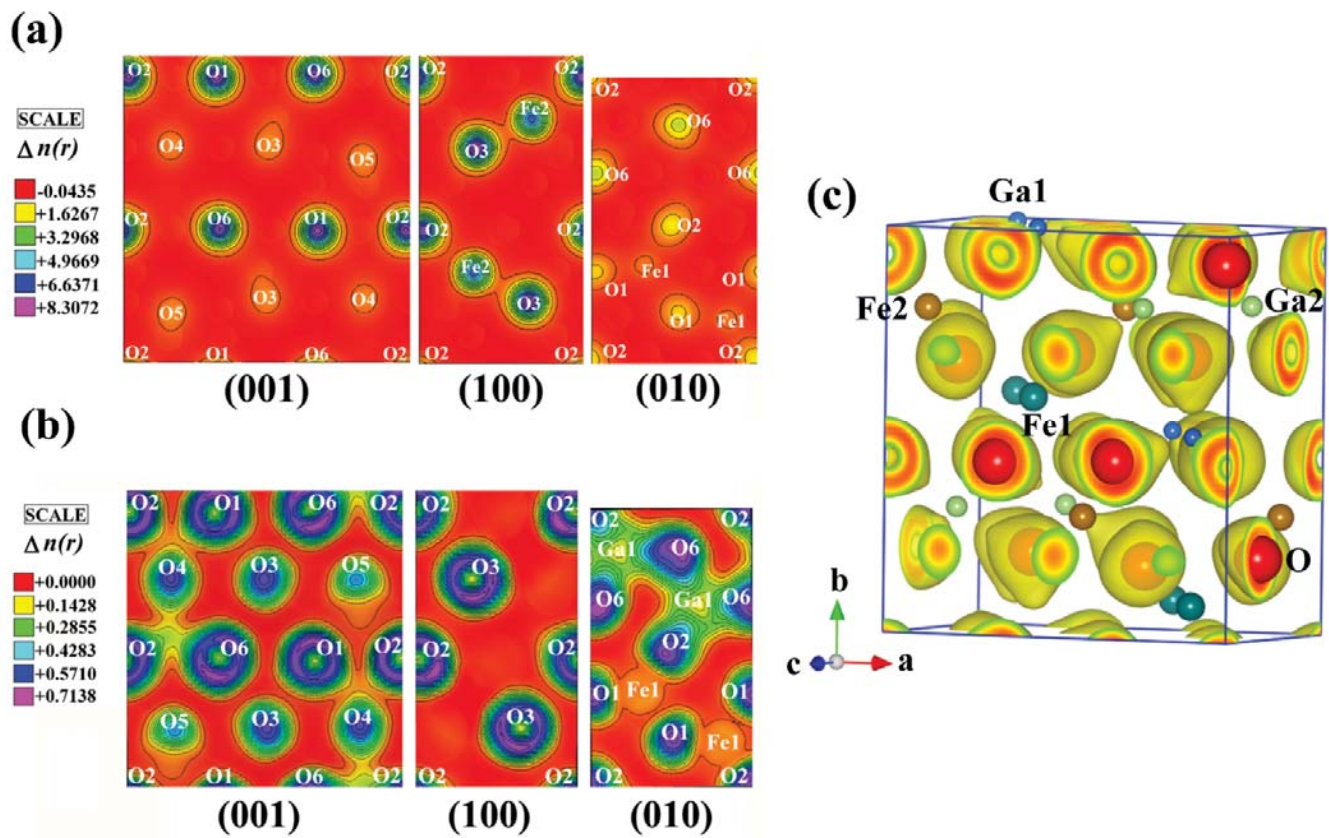

Fig.5. Roy *et al*,

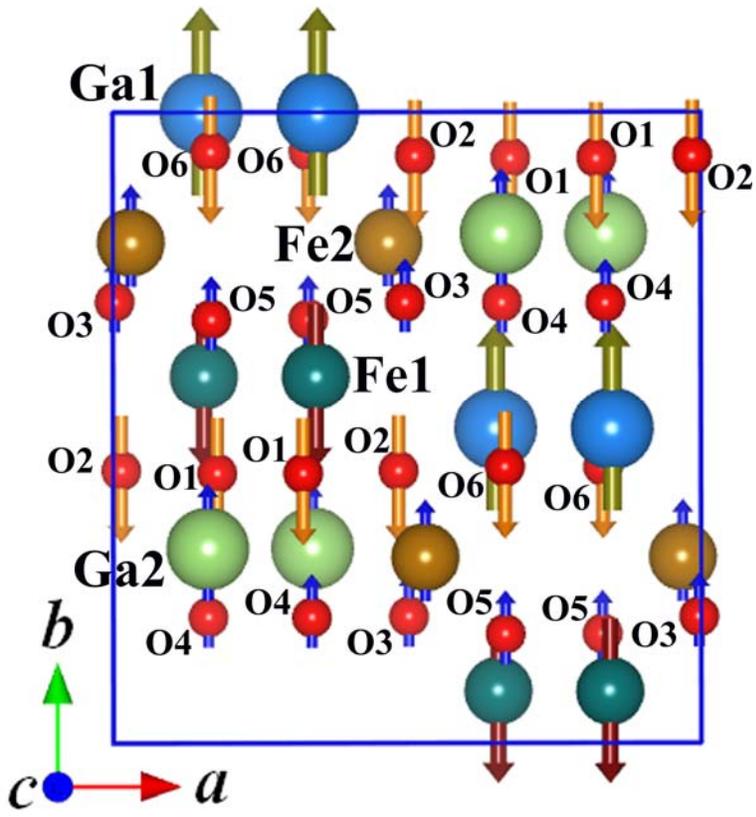

Fig.6. Roy *et al*,